\documentclass[aps,pra, 10pt,  twocolumn,amssymb,amsfonts,amsmath,compress,floatfix]{revtex4-1}

\usepackage[dvipdfmx]{graphicx}			% Graphics manager
\usepackage{subfig}
\usepackage{needspace} % Used to prevent some bad line breaks.
\usepackage{hyperref}
\hypersetup{pdfborder={0 0 0}, colorlinks, breaklinks=true,
  urlcolor=[rgb]{0,0.1,0.3}, citecolor=[rgb]{0,0.1,0.3}, linkcolor=[rgb]{0.0,0.1,0.3}
}

\begin{document}
%%%% Article title to be placed here
\title{Not Normal: the uncertainties of scientific measurements}

\newlength \figwidth % Define width of figures
\setlength \figwidth {1.0\linewidth}
\author{David C. Bailey}
\email{dbailey@physics.utoronto.ca}
\affiliation{Physics Department, University of Toronto, Toronto, ON, Canada M5S 1A7}
\date{\today}

\begin{abstract}
\normalsize
Judging the significance and reproducibility of quantitative research requires a good understanding of relevant uncertainties, but it is often unclear how well these have been evaluated and what they imply.
Reported scientific uncertainties were studied by analysing 41000 measurements of 3200 quantities from medicine, nuclear and particle physics, and interlaboratory comparisons ranging from chemistry to toxicology.
Outliers are common, with $5 \sigma$ disagreements up to five orders of magnitude more frequent than naively expected.
Uncertainty-normalized differences between multiple measurements of the same quantity are consistent with heavy-tailed \mbox{Student-t} distributions that are often almost Cauchy, far from a Gaussian Normal bell curve.
Medical research uncertainties are generally as well evaluated as those in physics, but physics uncertainty improves more rapidly, making feasible simple significance criteria such as the $5 \sigma$ discovery convention in particle physics.
Contributions to measurement uncertainty from mistakes and unknown problems are not completely unpredictable.
Such errors appear to have power-law distributions consistent with how designed complex systems fail, and how unknown systematic errors are constrained by researchers.
This better understanding may help improve analysis and meta-analysis of data, and help scientists and the public have more realistic expectations of what scientific results imply.
\end{abstract}

\maketitle
\section{Introduction}\label{sec:Introduction}
What do reported uncertainties actually tell us about the accuracy of scientific measurements and the likelihood that different measurements will disagree?
No scientist expects different research studies to always agree, but
the frequent failure of published research to be confirmed has generated much concern about scientific reproducibility \cite{McNutt2014,Conrad2015}.

When scientists investigate many quantities in very large amounts of data, interesting but ultimately false results may occur by chance and are often published.
In particle physics, bitter experience with frequent failures to confirm such results eventually led to an ad hoc ``5-sigma'' discovery criterion \cite{Rosenfeld1968,Franklin2013,Cousins2014,Dorigo2015}, i.e. a ``discovery'' is only taken seriously if the estimated probability for observing the result without new physics is less than the chance of a single sample from a Normal distribution being more than five standard deviations (``$5\sigma$'') from the mean.

In other fields, arguments that most novel discoveries are false \cite{Ioannidis2005} have caused increased emphasis on reporting the value and uncertainty of measured quantities, not just whether the value is statistically different from zero \cite{Nakagawa2007,Cumming2014}.
Research confirmation is then judged by how well independent studies agree according to their reported uncertainties, so assessing reproducibility requires accurate evaluation and realistic understanding of these uncertainties.
This understanding is also required when analysing data, combining studies in meta-analyses, or making scientific, business, or policy judgments based on research.
The experience of research fields such as physics, where values and uncertainties have long been regularly reported, may provide some guidance on what reproducibility can reasonably be expected \cite{Hedges1987}.

Most recent investigations into reproducibility focus on how often observed effects disappear in subsequent research, revealing strong selection bias in published results.
Removing such bias is extremely important, but may not reduce the absolute number of false discoveries since not publishing non-significant results does not make the ``discoveries'' go away.
Controlling the rate of false discoveries depends on establishing criteria that reflect real measurement uncertainties, especially the likelihood of extreme fluctuations and outliers \cite{Porter2006}.

Outliers are observations that disagree by an abnormal amount with other measurements of the same quantity.
Despite every scientist knowing that the rate of outliers is always greater than naively expected, there is no widely accepted heuristic for estimating the size or shape of these long tails.
These estimates are often assumed to be approximately Normal (Gaussian), but it is easy to find examples where this is clearly untrue \cite{Youden1972,Stigler1977,Speake2014}.

To examine the accuracy of reported uncertainties, this paper reviews multiple published measurements of many different quantities, looking at the differences between measurements of each quantity normalized by their reported uncertainties.
Previous similar studies \cite{Bukhvostov1973,Roos1975,Shlyakhter1994,Bukhvostov1997,Hanson2007,Jeng2007} reported on only a few hundred to a few thousand measurements, mostly in subatomic physics.
This study reports on how well multiple measurements of the same quantity agree, and hence what are reasonable expectations for the reproducibility of published scientific measurements.
Of particular interest is the frequency of large disagreements which usually reflect unexpected systematic effects.

\subsection{Systematic effects}
Sources of uncertainty are often categorized as statistical or systematic, and their methods of evaluation classified as Type A or B \cite{GUM2008}.
Type A evaluations are based on observed frequency distributions; Type B evaluations use other methods.
Statistical uncertainties are always evaluated from primary data using Type A methods, and can in principle be made arbitrarily small by repeated measurement or large enough sample size.

Uncertainties due to systematic effects may be evaluated by either Type A or B methods, and fall into several overlapping classes \cite{Sinervo2003}.
Class 1 systematics, which include many calibration and background uncertainties, are evaluated by Type A methods using ancillary data.
Class 2 systematics are almost everything else that might bias a measurement, and are caused by a lack of knowledge or uncertainty in the measurement model, such as the reading error of an instrument or the uncertainties in Monte Carlo estimates of corrections to the measurement.
Class 3 systematics are theoretical uncertainties in the interpretation of a measurement.
For example, determining the proton radius using the Lamb shift of muonic hydrogen requires over 20 theoretical corrections \cite{Carlson2015} that are potential sources of uncertainty in the proton radius, even if the actual measurement of the Lamb shift is perfect.
The uncertainties associated with Class 2 and 3 systematic effects cannot be made arbitrarily small by simply getting more data.

When considering the likelihood of extreme fluctuations in measurements, mistakes and ``unknown unknowns'' are particularly important, but they are usually assumed to be statistically intractable and are not often considered in traditional uncertainty analysis.
Mistakes are ``unknown knowns'', i.e. something that is thought to be known but is not, and it is believed that good scientists should not make mistakes.

``Unknown unknowns'' are factors that affect a measurement but are unknown and unanticipated based on past experience and knowledge \cite{Shields2015}.
For example, during the first 5 years of operation of LEP (the Large Electron Positron collider), the effect of local railway traffic on measurements of the $Z^0$ boson mass was an ``unknown unknown'' that no-one thought about. Then improved monitoring revealed unexpected variations in the accelerator magnetic field, and after much investigation these variations were found to be caused by electric rail line ground leakage currents flowing through the LEP vacuum pipe \cite{Bravin1998}.

In general, systematic effects are challenging to \mbox{estimate \cite{Youden1972,Colclough1987,Barlow2002,Sinervo2003,GUM2008,Pavese2009b,Attivissimo2011}},
but can be partially constrained by researchers making multiple internal and external consistency checks:
Is the result compatible with previous data or theoretical expectations?
Is the same result obtained for different times, places, assumptions, instruments, or subgroups?
As described by Dorsey  \cite{Dorsey1944}, scientists ``change every condition that seems by any chance likely to affect the result, and some that do not, in every case pushing the change well beyond any that seems at all likely''.
If an inconsistency is observed and its cause understood, the problem can often be fixed and new data taken, or the effect monitored and corrections made.
If the cause cannot be identified, however, then the observed dispersion of values must be included in the uncertainty.

The existence of unknown systematic effects or mistakes may be revealed by consistency checks \cite{Pomme2016}, but small unknown systematics and mistakes are unlikely to be noticed if they do not affect the measurement by more than the expected uncertainty.
Even large problems can be missed by chance (see Sec.~\ref{ssec:Modelling}) or if the conditions changed between consistency checks do not alter the size of the systematic effect.
The power of consistency checks is limited by the impossibility of completely changing all apparatus, methods, theory, and researchers between measurements, so one can never be certain that all significant systematic effects have been identified.

\section{Methods}\label{sec:Methods}

\subsection{Data}
Quantities were only included in this study if they are significant enough to have warranted at least five independent measurements with clearly stated uncertainties.

Medical and health research data were extracted from some of the many meta-analyses published by the Cochrane collaboration \cite{Cochrane};
a total of 5580 measurements of 310 quantities generating 99433 comparison pairs were included.
Particle physics data (8469 measurements, 864 quantities, 53988 pairs) were retrieved from the Review of Particle Physics~\cite{RPP2012,RPP2013}.
Nuclear physics data (12380 measurements, 1437 quantities, 66677 pairs) were obtained from the Table of Radionuclides \cite{BIPM-5}.

Most nuclear and particle physics measurements have prior experimental or theoretical expectations which may influence results from nominally independent experiments \cite{Jeng2007,Baker2013}, and medical research has similar biases \cite{Ioannidis2005}, so this study also includes a large sample of interlaboratory studies that do not have precise prior expectations for their results.
In these studies, multiple independent laboratories measure the same quantity and compare results.
For example, the same mass standard might be measured by national laboratories in different countries, or an unknown archaeological sample might be divided and distributed to many labs, with each lab reporting back its Carbon-14 measurement of the sample's age.
None of the laboratories knows the expected value for the quantity nor the results from other labs, so there should be no expectation, selection, or publication biases.
These Interlab studies (14097 measurements, 617 quantities,  965416 pairs) were selected from a wide range of sources in fields such as analytical chemistry, environmental sciences, metrology, and toxicology. The measurements ranged from genetic contamination of food to high precision comparison of fundamental physical standards, and were carried out by a mix of national, university, and commercial laboratories.

All quantities analysed are listed in the Supplementary Materials \cite{Supplementary}.

\subsection{Data selection and collection}
Data were entered using a variety of semi-automatic scripts,  optical-character recognition, and manual methods.
No attempt was made to recalculate past results based on current knowledge, or to remove results that were later retracted or amended, since the original paper was the best result at the time it was published.
When the Review of Particle Physics \cite{RPP2013} noted that earlier data had been dropped, the missing results were retrieved from previous editions~\cite{RPParchive}.

To ensure that measurements were as independent as possible, measurements were excluded if they were obviously not independent of other data already included.
Because relationships between measurements are often obscure, however, there undoubtedly remain many correlations between the published results used.

Medical and health data were selected from the 8105 reviews in the Cochrane database \cite{Cochrane} as of 25 September 2013. Data were analysed from 221 Intervention Reviews whose abstract mentioned $\ge6$ trials with $\ge5000$ total participants, and which reported at least one analysis with $\ge5$ studies with $\ge3500$ total participants.
The average heterogeneity inconsistency index ($I^2\equiv1-\textrm{dof/}\chi^2$) \cite{Higgins2003,Baker2013} is about 40\% for the analyses reported here.
Because analyses within a review may be correlated, only a maximum of 3 analyses and 5 comparison groups were included from any one review.
About 80\% of the Cochrane results are the ratio of intervention and control binomial probabilities, e.g. mortality rates for a drug and a placebo.
Such ratios are not Normal~\cite{Andres2014}, so they were converted to differences that should be Normal in the Gaussian limit, i.e. when the group size $n$ and probability $p$ are such that $n$, $np$, and $(1-p)n$ are all  $>>1$, so the binomial distribution converges towards a Gaussian distribution.
(The median observed values for these data were $n=100$, $p=0.16$.)
The 68.3\% binomial probability confidence interval was calculated for both the intervention and control groups to determine the uncertainties.

\subsection{Uncertainty evaluation}
Measurements with uncertainties are typically reported as ${x \pm u}$, which means that the interval ${x - u}$ to ${x + u}$ contains with some defined probability ``the values that could reasonably be attributed to the measurand'' \cite{GUM2008}.
Most frequently, uncertainty intervals are given as $\pm k u_S$, where $k$ is the coverage factor and $u_S$ is the ``standard uncertainty'', i.e. the uncertainty of a measurement expressed as the standard deviation of the expected dispersion of values.
Uncertainties in particle physics and medicine are often instead reported as the bounds of either 68.3\% or 95\% confidence intervals, which for a Normal distribution are equivalent to the $k=1$ and $2$ standard uncertainty intervals.

For this study, all uncertainties were converted to nominal 68.3\% confidence interval uncertainties.
The vast majority of measurements reported simple single uncertainties, but if more than a single uncertainty was reported, e.g. ``statistical'' and ``systematic'', they were added in quadrature.

\subsection{Normalized differences}
All measurements, ${x_i \pm u_i}$, of a given quantity were combined in all possible pairs and the difference between the two measurements of each pair calculated in units of their combined uncertainty $u_{ij}$:
\begin{equation}
\label{eq:zDefinition}
z_{ij}=\frac{|x_i - x_j|}{\sqrt{u_i^2+u_j^2}}.
\end{equation}
The dispersion of $z_{ij}$ values can be used to judge whether independent measurements of a quantity are ``compatible'' \cite{VIM2012}.
A feature of $z$ as a metric for measurement agreement is that it does not require a reference value for the quantity. (The challenges and effects of using reference values are discussed in Section \ref{ssec:Alternate}.)

The uncertainties in Equation~\ref{eq:zDefinition} are combined in quadrature, as expected for standard uncertainties of independent measurements.
(The effects of any lack of independence are discussed in Section~\ref{ssec:UncertaintySchemes}.)

Uncertainties based on confidence intervals may not be symmetric about the reported value, which is the case for about 13\% of Particle, 6\% of Medical, 0.3\% of Nuclear, and 0.06\% of Interlab measurements.
Following common (albeit imperfect) practice \cite{Barlow2003}, if the reported plus and minus uncertainties were asymmetric, $z_{ij}$ was calculated from Eq.~\ref{eq:zDefinition} using the uncertainty for the side towards the other member of the comparison pair.
For example, if $x_1 = 80 \pm ^{3}_{2}$, $x_2 = 100 \pm ^{5}_{4}$, and  $x_3 = 126 \pm ^{15}_{12}$, then $z_{12}={(100-80)}/\sqrt{3^2+4^2}$ and $z_{23}=(126-100)/\sqrt{5^2+12^2}$.

\begin{figure*}
\centering
\subfloat{
  \includegraphics[width=69mm]{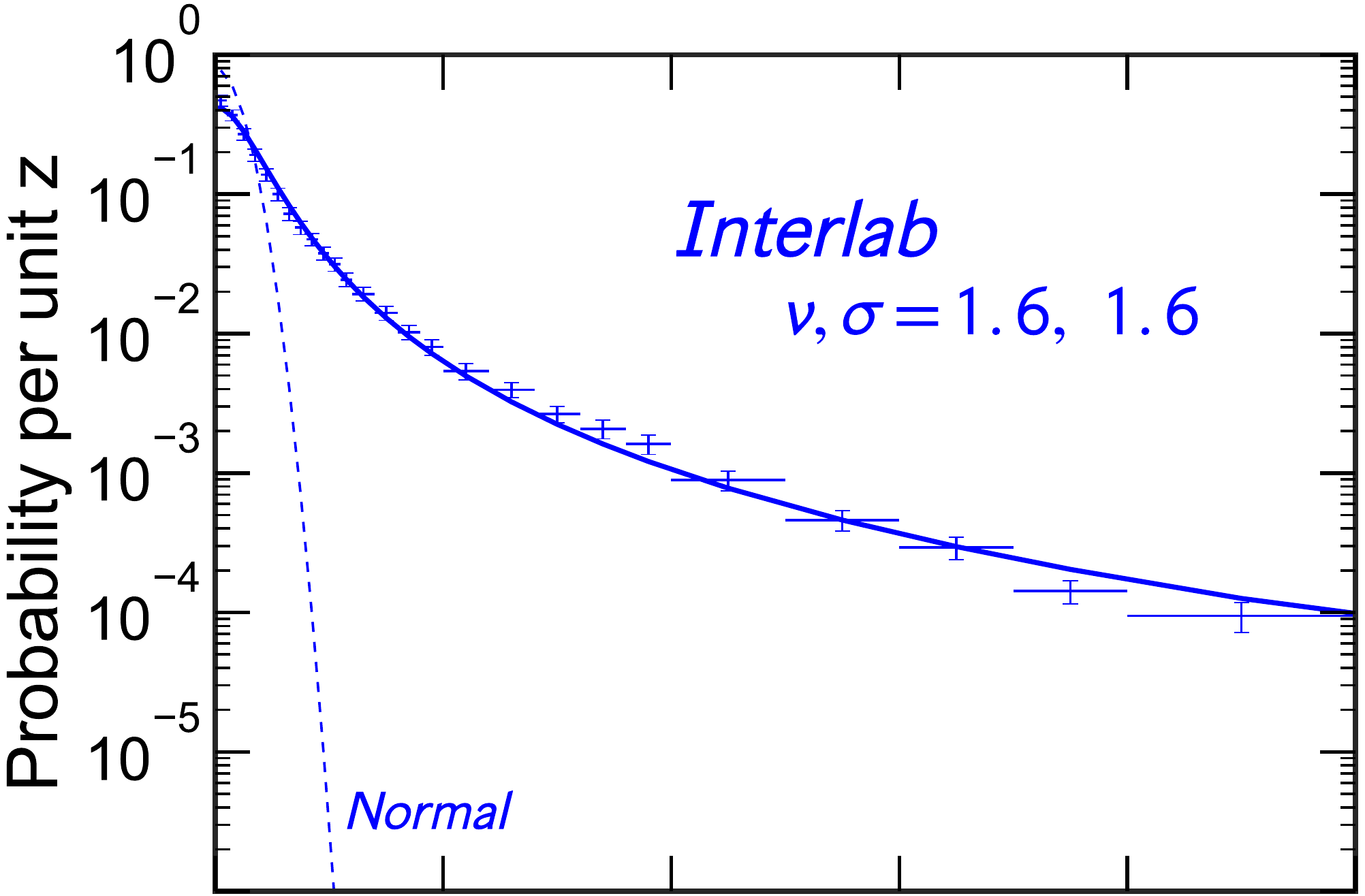}\quad
}
\subfloat{
  \includegraphics[width=60mm]{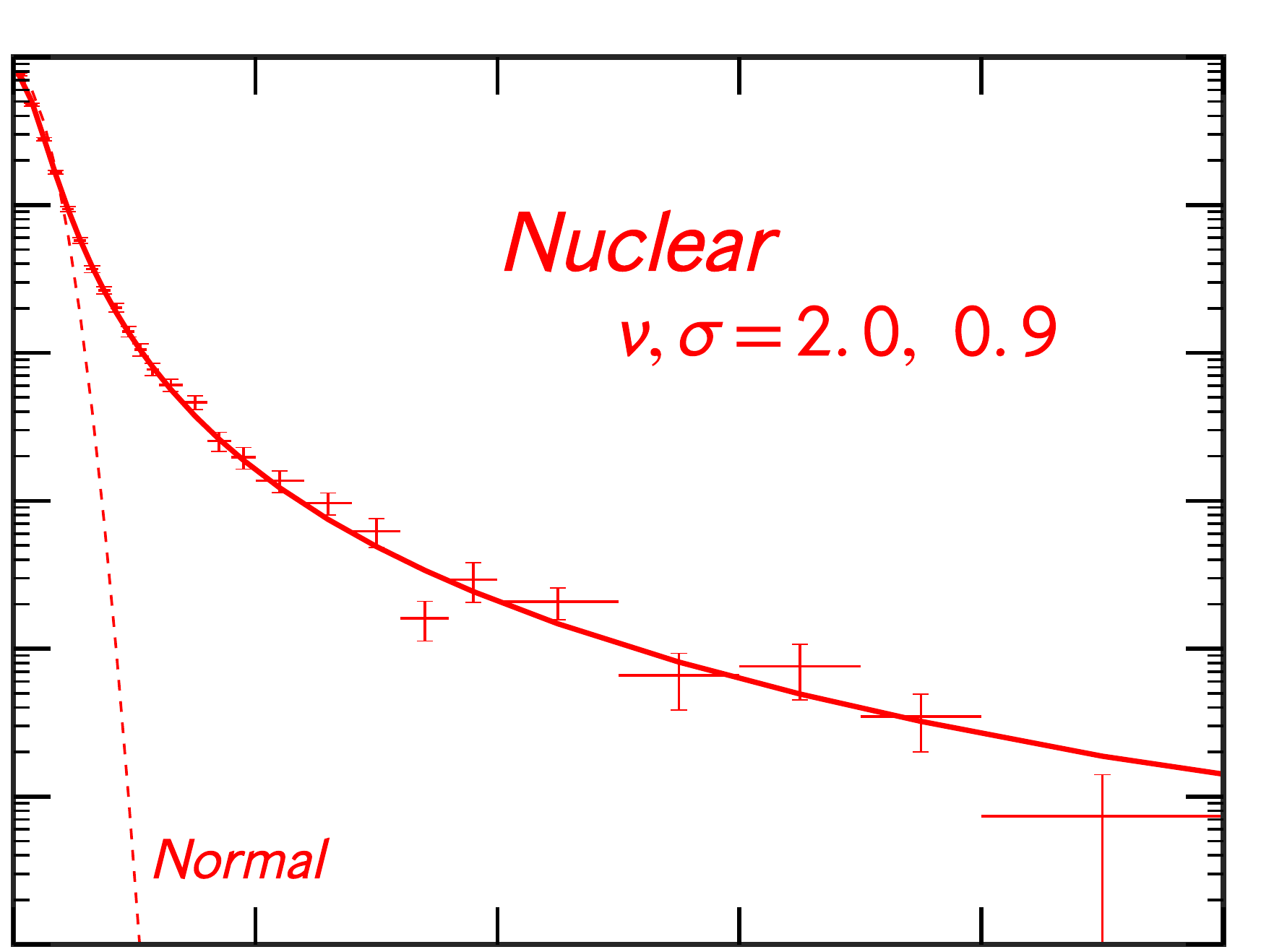}
}
\hspace{0mm}
\subfloat{
  \includegraphics[width=71mm]{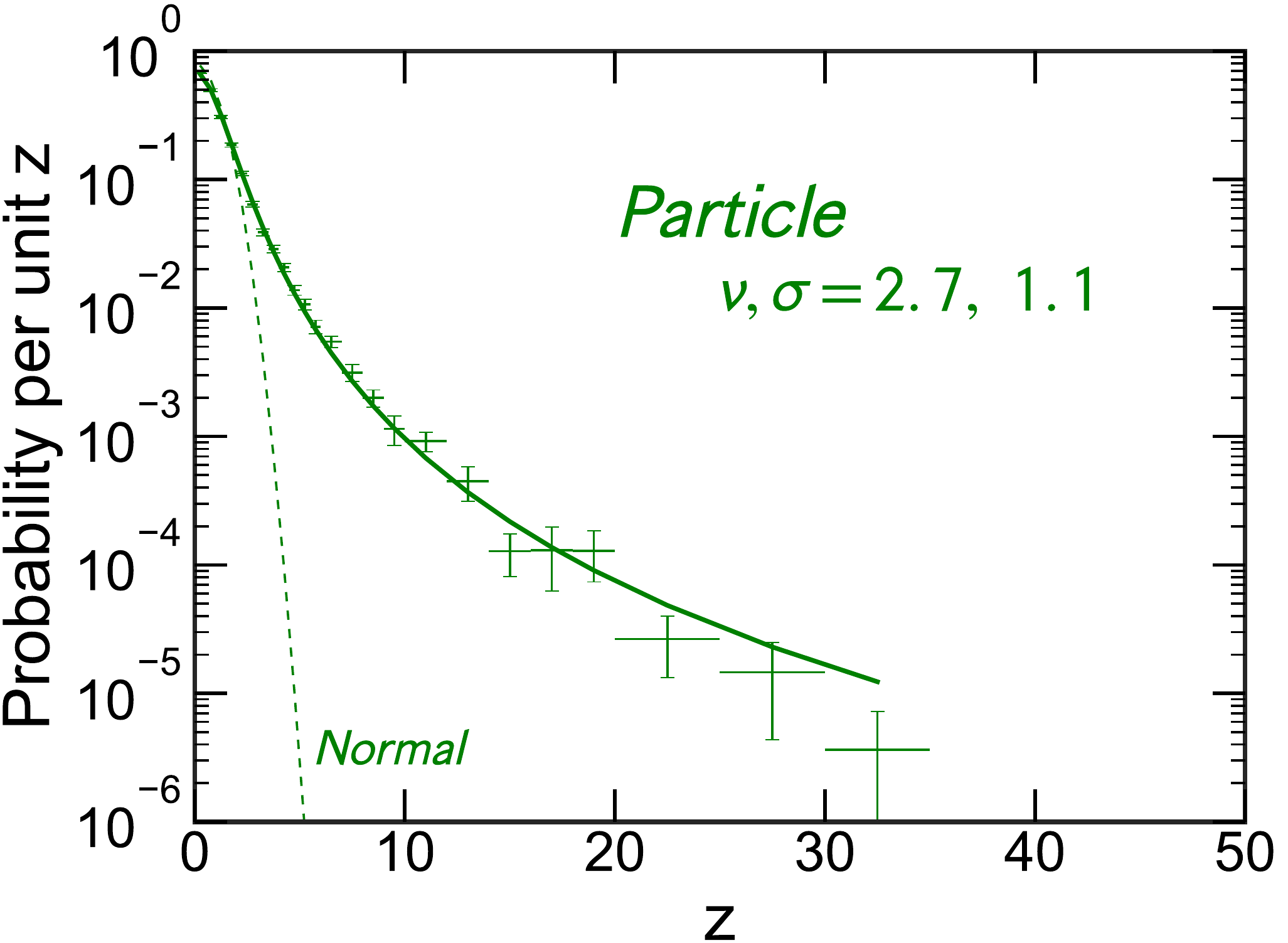}
}
\subfloat{
  \includegraphics[width=60.5mm]{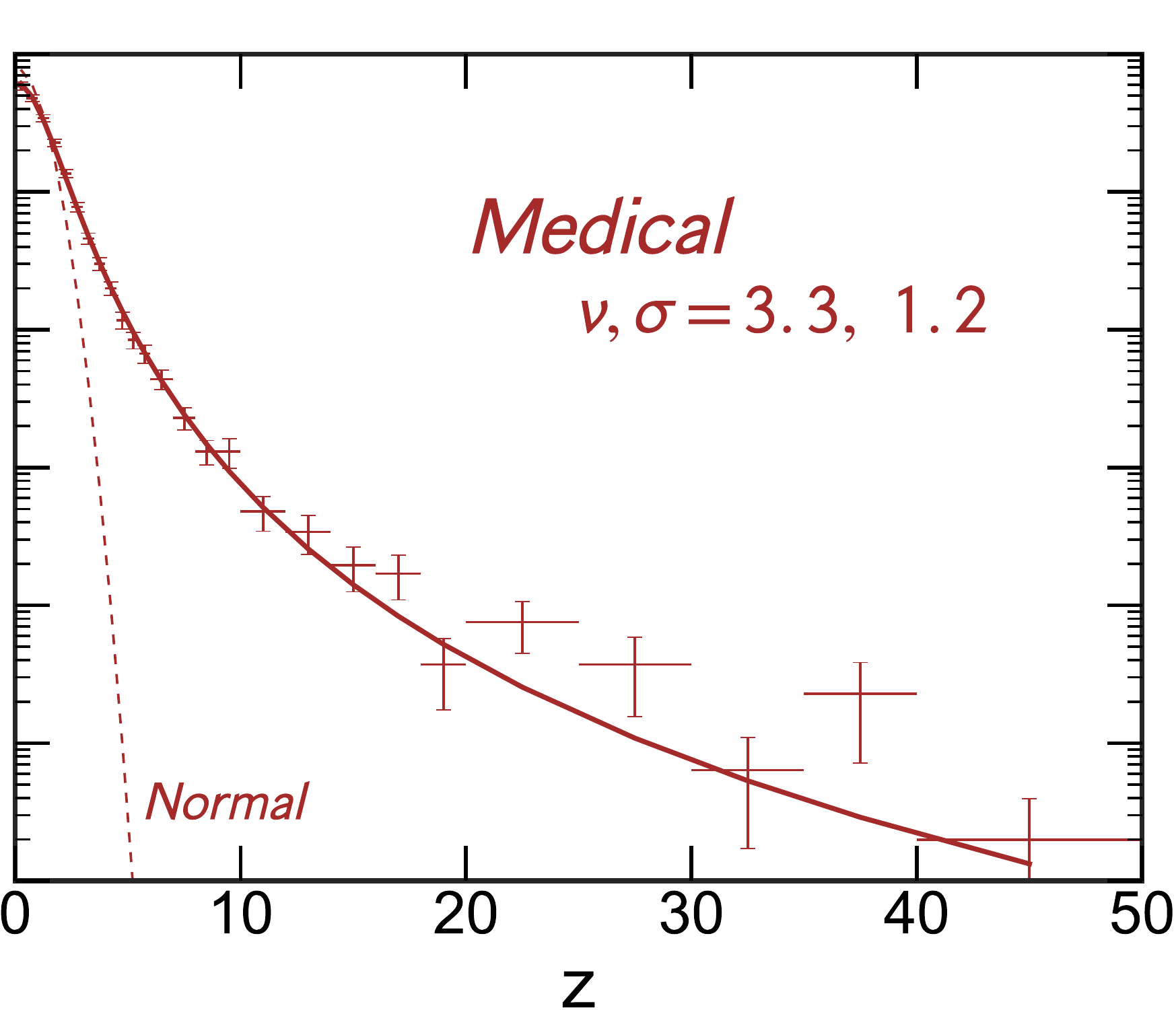}
}
\caption{\label{fig:differential}Histograms of uncertainty normalized differences ($z_{ij}$ from Eq.~\ref{eq:zDefinition}) per unit of z.  Horizontal and vertical error bars are the bin width and the standard uncertainty evaluated by a bootstrap Monte Carlo. The smooth curves are best-fit Student-t distributions. The dashed curves are Normal distributions.}
\end{figure*}

The distributions of the $z_{ij}$ differences are histogrammed in Fig.~\ref{fig:differential}, with each pair weighted such that the total weight for a quantity is the number of measurements of that quantity.
For example, if a quantity has 10 measurements, there are 45 possible pairs, and each entry has a weight of 10/45.
(Other weighting schemes are discussed in Section \ref{ssec:Alternate}.)
The final frequency distribution within each research area is then normalized so that its total observed probability adds up to 1.
If the measurement uncertainties are well evaluated and correspond to Normally distributed probabilities for $x$, then $z$ is expected to be Normally distributed with a standard deviation $\sigma=1$.

Probability distribution uncertainties (e.g. the vertical error bars in Fig.~\ref{fig:differential}) were evaluated using a bootstrap Monte Carlo method where quantities were drawn randomly with replacement from the actual data set until the number of Monte Carlo quantities equaled the actual number of quantities.
The resulting artificial data set was histogrammed, the process repeated 1000 times, and the standard deviations of the Monte Carlo probabilities calculated for each $z$ bin.

Random selection of measurements instead of quantities was not chosen for uncertainty evaluation because of the corrections then required to avoid bias and artifacts.
For example, if measurements are randomly drawn, a quantity with only 5 measurements will often be missing from the artificial data set for having too few ($<5$) measurements drawn, or if it does have 5 measurements some of them will be duplicates generating unrealistic $z=0$ values, or if duplicates are excluded then they will always be the same 5 nonrandom measurements.
Without correcting for such effects, the resulting measurement Monte Carlo generated uncertainties are too small to be consistent with the observed bin-to-bin fluctuations in Fig.~\ref{fig:differential}.
Correcting for such effects requires using characteristics of the actual quantities and would be effectively equivalent to using random quantities.

\subsection{Data fits}
Attempts were made to fit the data to a wide variety of functions, but by far the best fits were to non-standardized Student-t probability density distributions with $\nu$ degrees of freedom.
\begin{equation}
\label{eq:Snuz}
S_{\nu,\sigma}(z)=\frac{\Gamma\left((\nu+1)/2\right)}{\Gamma\left(\nu/2\right)}\frac{1}{\sqrt{\nu\pi}\sigma}\frac{1}{\left(1+(z/\sigma)^2/\nu\right)^{(\nu+1)/2}}
\end{equation}
A Student-t distribution is essentially a smoothly symmetric normalizable power-law, with $S_{\nu,\sigma}(z)\sim(z/\sigma)^{-(\nu+1)}$ for $|z|\gg \sigma\sqrt{\nu}$.

The fitted parameter $\sigma$  defines the core width and overall scale of the distribution and is equal to the standard deviation in the \mbox{$\nu\rightarrow\infty$} Gaussian limit and to the half-width at half maximum in the \mbox{$\nu \rightarrow 1$} Cauchy (also known as Lorentzian or Breit-Wigner) limit.
The parameter $\nu$ determines the size of the tails, with small $\nu$ corresponding to large tails.
The values and standard uncertainties in $\sigma$ and $\nu$ were determined from a non-linear least squares fit to the data that minimizes the nominal $\chi^2$ \cite{BohmZech2010}:
\begin{equation}
\chi^2  = \sum\limits_{i=1}^{N_{bins}} \frac{\left (B_i - S_{\nu,\sigma}(z_i) \right )^2}{u_{B_i}^2}
\end{equation}
where $z_i$, $B_i$ and $u_{B_i}$ are the bin $z$, contents, and uncertainties of the observed $z$ distributions shown in Fig.~\ref{fig:differential}.

Possible values of $z$ are sometimes limited by the allowed range of measurement values, which could suppress heavy tails.
For example, many quantities are fractions that must lie between 0 and 1, and there is less room for two measurements with 10\% uncertainty to disagree by $5 \sigma$ than for two 0.01\% measurements.
The size of this effect was estimated using Monte Carlo methods to generate simulated data based on the values and uncertainties of the actual data, constrained by any obvious bounds on their allowed values.
The simulated data was then fit to see if applying the bounds changed the fitted values for $\sigma$ and $\nu$.
The largest effect was for Medical data where $\nu$ was reduced by about 0.1 when minimally restrictive bounds were assumed.
Stronger bounds might exist for some quantities, but determining them would require careful measurement-by-measurement assessment beyond the scope of this study.
For example, each measurement of the long-term duration of the effect of a medical drug or treatment would have an upper bound set by the length of that study.
Since correcting for bounds can only make $\nu$ smaller (corresponding to even heavier tails), and the observed effects were negligible, no corrections were applied to the values of $\nu$ reported here.

\Needspace{6\baselineskip} % Prevent bad page break here

\section{Results}\label{sec:Results}

\subsection{Observed distributions}

\begin{figure}[ht]
\centering
\includegraphics[width=\figwidth]{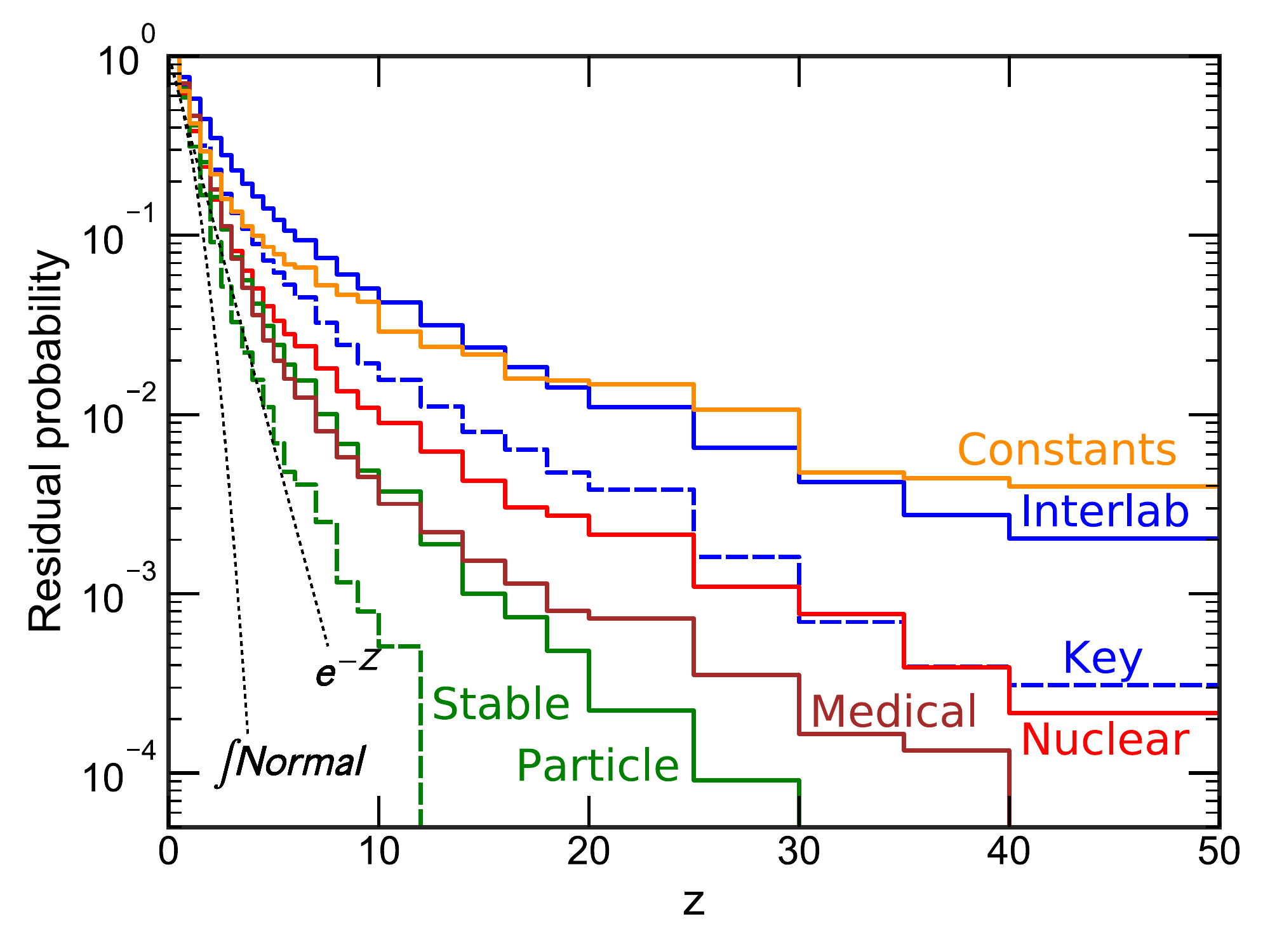}
\caption{\label{fig:cumulative}The observed probability of two measurements disagreeing by more than $z$ standard uncertainties for different data sets:  $\int_z^\infty \mathcal{P}(x) dx$. (See also Table~\ref{tab:CL})}
\end{figure}

Histograms of the $z$ distributions for different data sets are shown in Fig.~\ref{fig:differential}.
The complementary cumulative distributions of the data are given in Table~\ref{tab:CL} and shown in Fig.~\ref{fig:cumulative}.

\begin{table*}[ht]
\newcommand{\DCBindent}{\hspace{5mm}}
\centering
\caption{\label{tab:CL}Observed chance of experimental disagreement by more than $z$ standard uncertainties for different data sets, compared to values expected for some theoretical distributions. Also listed are the $z$ values that bound 95\% of the distribution, i.e. $p_{true}=0.05$, and $p$ values for that $z$ for a Normal distribution.}
\medskip
\begin{tabular*}{\hsize}{@{\extracolsep{\fill}}llllll|rr}
\hline
~~~~~~~~~~~~~~~~$z>	$				 & 1	 & 2  	  & 3 		& 5					& 10				 & $z_{0.95}$&$p_{Normal}(z_{0.95})$\\
\hline
\rule{0pt}{1em}Interlab				 & $0.58$& $0.35$ & $0.23$ 	& $0.12$			& $0.042$			 & $9.0$	 & $2\times10^{-19}$	  \\
\DCBindent(Key)						 & $0.46$& $0.23$ & $0.13$	& $0.062$			& $0.016$			 & $5.7$	 & $1\times10^{-8}$	  	  \\
Nuclear 							 & $0.38$& $0.16$ & $0.082$	& $0.033$			& $0.009$			 & $4.0$	 & $6\times10^{-5}$  	  \\
Particle							 & $0.41$& $0.16$ & $0.075$	& $0.024$			& $0.004$			 & $3.7$	 & $2\times10^{-4}$  	  \\
\DCBindent(Stable)					 & $0.31$& $0.091$& $0.033$ & $0.007$			& $0.0005$			 & $2.5$	 & $1\times10^{-2}$	      \\
Medical								 & $0.47$& $0.18$ & $0.074$	& $0.020$			& $0.003$			 & $3.5$	 & $4\times10^{-4}$  	  \\
Constants							 & $0.42$& $0.22$ & $0.14$	& $0.078$			& $0.029$			 & $7.2$	 & $6\times10^{-13}$ 	  \\
\hline
\rule{0pt}{1em}Normal (Gaussian)	 & $0.32$& $0.046$& $0.0027$& $5.7\times10^{-7}$& $1.5\times10^{-23}$& $1.96$	 & $5\times10^{-2}$  	  \\
Student-t ($\nu=10$)				 & $0.34$& $0.073$& $0.013$& $5.4\times10^{-4}$& $1.6\times10^{-6}$ & $2.23$	 & $2.6\times10^{-2}$ 	  \\
Exponential							 & $0.37$& $0.14$ & $0.050$	& $0.007$			& $4.5\times10^{-5}$ & $3.0$	 & $2.7\times10^{-3}$	  \\
Student-t ($\nu=2$)					 & $0.42$& $0.18$ & $0.095$	& $0.038$			& $0.010$			 & $4.3$	 & $2\times10^{-15}$ 	  \\
Cauchy								 & $0.50$& $0.30$ & $0.20$	& $0.13$			& $0.063$			 & $12.8$	 & $2\times10^{-37}$ 	  \\
\hline
\end{tabular*}
\end{table*}

None of the data are close to Gaussian, but all can reasonably be described by almost-Cauchy Student-t distributions with $\nu \sim 2\!-\!3$.
For comparison, fits to these data with L\'{e}vy stable distributions have nominal $\chi^2$  4 to 30 times worse than the fits to Student-t distributions.
The number of ``$5 \sigma$'' (i.e. $z>5$) disagreements observed is as high as $0.12$, compared to the $6 \times 10^{-7}$ expected for a Normal distribution.

The fitted values for $\nu$ and $\sigma$ are shown in Table~\ref{tab:Fits}.
Also shown in Table~\ref{tab:Fits} are two data subsets expected to be of higher quality, BIPM Interlaboratory Key comparisons (372 quantities, 3712 measurements, 20245 pairs) and Stable Particle properties (335 quantities, 3041 measurements, 16649 pairs).
The Key comparisons \cite{BIPMKey2014} should define state-of-the-art accuracy, since they are measurements of important metrological standards carried out by national laboratories.
Stable particles are often easier to study than other particles, so their properties are expected to be better determined.
Both ``better'' data subsets do have narrower distributions consistent with higher quality, but they still have heavy tails.
More selected data subsets are discussed in Section \ref{ssec:Selected subsets}.

\renewcommand{\tabcolsep}{2pt}

\begin{table*}[ht]
\newcommand{\DCBindent}{\hspace{5mm}}
\centering
\caption{\label{tab:Fits}Fitted Student-t parameters with nominal $\chi^2$ per degree-of-freedom. Also shown are parameters for quantities with ${\ge}10\;$
measurements, for newer measurements made since the year 2000, and for the approximate distribution of individual measurements.
Uncertainties not shown for $\sigma_{10}$, $\sigma_{new}$, $\nu_x$ and $\sigma_x$  are  ${\lesssim}0.1$.}
\medskip
\begin{tabular}{l||c|c|c||c|c||c|c||cc}
\hline
\multicolumn{1}{l||}{} &
	\multicolumn{1}{c}{$\nu$} &
		\multicolumn{1}{c}{$\sigma$} &
			\multicolumn{1}{c||}{$\chi^2\textrm{/dof}$} &
				\multicolumn{1}{c}{$\nu_{10}$} &
					\multicolumn{1}{c||}{$\ \sigma_{10}\ $} &
						\multicolumn{1}{c}{$\nu_{new}$} &
							\multicolumn{1}{c||}{$\ \sigma_{new}\ $} &
								\multicolumn{1}{c}{$\nu_x$} &
									\multicolumn{1}{c}{$\sigma_x$}\\
\hline
Interlab 			&$1.64\pm0.05$	&$1.62\pm0.05$	&$1.1$	&$1.6\pm0.1$	&$1.7      $	&$1.6\pm0.1$ &$1.7      $	&$1.5$ 	& $1.3$		\\
\DCBindent Key		&$1.90\pm0.10$	&$1.12\pm0.04$	&$1.9$	&$1.7\pm0.1$	&$1.1      $	&$1.9\pm0.1$ &$1.2      $	&$1.7$	& $0.9$		\\
Nuclear				&$1.99\pm0.06$	&$0.90\pm0.02$	&$1.6$	&$2.4\pm0.2$	&$1.1      $	&$2.1\pm0.2$ &$0.9      $	&$1.8$ 	& $0.7$ 	\\
Particle			&$2.75\pm0.10$	&$1.05\pm0.02$	&$1.5$	&$2.8\pm0.1$	&$1.1      $	&$2.6\pm0.2$ &$1.0      $	&$2.4$	& $0.9$		\\
\DCBindent Stable	&$3.45\pm0.16$	&$0.86\pm0.02$	&$0.6$	&$3.8\pm0.4$	&$0.9      $	&$7.6\pm1.3$ &$0.9      $	&$2.9$	& $0.8$ 	\\
Medical				&$3.30\pm0.11$	&$1.18\pm0.02$	&$0.7$	&$3.3\pm0.1$	&$1.2      $	&$3.2\pm0.2$ &$1.2      $	&$2.8$ 	& $1.0$		\\
Constants			&$1.81\pm0.15$	&$0.89\pm0.06$	&$0.8$	&$1.8\pm0.2$	&$0.9      $	&$1.3\pm0.3$ &$1.1      $	&$1.7$ 	& $0.7$		\\
\hline
Normal				&$\infty$		&$1.0$			&		&   	   		&				&			 &   			&$\infty$ & $1.0$	\\
Cauchy				&$1.0$			&$\sqrt{2}^*$	&		&   	   		&				&			 &   			&$1.0$	& $1.0$		\\
\hline
\multicolumn{2}{l}{}&\multicolumn{7}{l}{$^*$ For uncertainties added in quadrature} \\

\end{tabular}
\end{table*}

The probability distribution for the nominal $\chi^2$ statistic is not expected to be an exact regular $\chi^2$ distribution.
The differences are due to the non-Gaussian uncertainties of the low-population high-$z$ bins, and because the bin contents are not independent since a single measurement can contribute to multiple bins as part of different permutation pairs.
Based on fits of simulated data sets with a mix of $\nu$ comparable to the observed data, the range of nominal $\chi^2$ reported in Table~\ref{tab:Fits} seems reasonable, i.e. the chances of $\chi^2\textrm{/dof}$ $\le0.6$ or $\ge1.9$ were $15\%$ and $2\%$ respectively.

To see if more important quantities are measured with less disagreement, a small additional data set of measurements of fundamental physical constants (7 quantities, 320 measurements, 9098 pairs) was also analysed.
The constants are
Avogadro's number, the fine structure constant, the Planck constant, Newton's gravitational constant, the deuteron binding energy, the Rydberg constant, and the speed of light (before it became a defined constant).
These measurements have very heavy tails, despite their importance in physical science.
Quantities with more interest do not seem to be better measured, as is also shown by considering only quantities with at least 10 published measurements, which do not have significantly smaller tails (see $\sigma_{10}$,$\nu_{10}$ in Table~\ref{tab:Fits}).

Fig.~\ref{fig:differential} shows that the comparison pairs $z_{ij}$ are Student-t distributed, but what does this imply about the dispersion of individual $x_i$ measurements?
Except for the $\nu=1$ and $\infty$ Cauchy and Normal limits, the distribution of differences of values selected from a Student-t distribution is not itself a t distribution, but it can be closely approximated as one \cite{Willink2004}.
The distributions of the parent individual $x$ measurements were estimated by Monte Carlo deconvolution.
Artificial measurements were generated from t distributions with parameters $\nu_x$ and $\sigma_x$, and these measurements combined into permutation pairs to generate an artificial $z$ distribution. This distribution was compared to the observed $z$ distributions, and then $\nu_x$ and $\sigma_x$ were iteratively adjusted until the best match was achieved between the artificial and observed $z$ distributions.
As shown in Table~\ref{tab:Fits}, the approximate Student-t parameters ($\nu_x$, $\sigma_x$) of the individual measurement populations have $\nu_x\!\!<\!\nu$ and hence are slightly more Cauchy-like than the permutation pairs distributions.

\subsection{\label{ssec:UncertaintySchemes}Combined uncertainty}
The definition of $z$ by  Equation~\ref{eq:zDefinition} assumes that the measurements $x_i$ and $x_j$ are independent and that the uncertainties $u_i$ and $u_j$ can be combined following the rules for standard  uncertainties.

If $x_i$ and $x_j$ are correlated, however, Equation~\ref{eq:zDefinition} should be replaced by
\begin{equation}
\label{eq:zcovDefinition}
z_{ij}=\frac{|x_i - x_j|}{\sqrt{u_i^2-2\textrm{cov}(x_i,x_j)+u_j^2}}.
\end{equation}
where $\textrm{cov}(x_i,x_j)$ is the covariance of $x_i$ and $x_j$ \cite{GUM2008}.

It is not in general possible to quantitatively evaluate the covariance for individual pairs of measurements in the data sets, but the effects of any correlations are not expected to be large, and they cannot explain the observed heavy tails.
Any positive covariance would decrease the denominator in Eq.~\ref{eq:zcovDefinition} and increase the width of the $z$ distributions.
Correlations between measurements are expected to be much more likely positive than negative, but even perfect anti-correlation could only decrease $z$ values by at most a factor of $1/\sqrt{2}$ compared to the uncorrelated case.
(i.e. Changing $\textrm{cov}(x_i,x_j)$ from $0$ to $-u_iu_j$ in Eq.~\ref{eq:zcovDefinition}  reduces $z_{ij}$ by $\sqrt{2}$ if $u_i=u_j$, and less if $u_i \ne u_j$.)
Correlations are further discussed in Section~\ref{ssec:Expectations}.

Another possible issue with Equation~\ref{eq:zDefinition} is that its usual derivation assumes that $u_i$ and $u_j$ are standard deviations of the expected dispersion of possible values (e.g. see Sec. E.3.1 of Ref.~\cite{GUM2008}).
This assumption is a concern since the standard deviation is an undefined quantity for Student-t distributions if $\nu<2$, and the observed $z$ and inferred $x$ distributions have $\nu$ near or below this value.
Even if the variance of a distribution is undefined, however, the dispersion of the difference of two independent variables drawn from such distributions may still be calculated numerically and in some cases analytically.

Cauchy uncertainties add linearly instead of in quadrature, since the distribution of differences of two variables drawn from two Cauchy distributions with widths $\sigma_1$ and $\sigma_2$ is simply another Cauchy distribution with width $\sigma_{diff}=\sigma_1+\sigma_2$.
The corresponding definition of $z$ would be
\begin{equation}
\label{eq:zCauchy}
z_{ij}^{Cauchy}=\frac{|x_i - x_j|}{u_i+u_j}.
\end{equation}
Almost-Cauchy distributions should almost follow the rules for combination of Cauchy  ($\nu=1$) distributions.
Applying Equation~\ref{eq:zCauchy} to the data produces $z$ distributions that appear almost identical to those in Fig.~\ref{fig:differential}, except that the fitted values of $\sigma$ for Interlab, Nuclear, Particle, Medical data are smaller by factors of $0.78,0.80,0.75,0.74$, while the the fitted values of $\nu$ are almost unchanged ($\nu_{linear}/\nu_{quad} = 0.99, 1.00, 0.98, 0.94$).
The scale factor for $\sigma$ would be $1/\sqrt{2}=0.71$ if all measurements of a quantity had equal uncertainties ($u_i=u_j$), since switching from quadrature (Eq.~\ref{eq:zDefinition}) to linear (Eq.~\ref{eq:zCauchy}) would simply scale all the calculated $z$ values by $1/\sqrt{2}$ and not affect $\nu$.
Similarly, if data with equal $\nu=1,\sigma=1$ Cauchy uncertainties were analysed using Equation~\ref{eq:zDefinition}, the resulting permutation pairs would have $\nu=1,\sigma=\sqrt{2}$, as shown in the last line of Table~\ref{tab:Fits}.

\renewcommand{\tabcolsep}{4pt}

\subsection{\label{ssec:Alternate}Alternate weighting schemes and compatibility measures}

There are several ways to weight data in the distribution plots, but the fitted parameter values are not usually greatly affected by the choice (see Table~\ref{tab:AlternateFits}).
The default method was to give each measurement equal weight (``M'' in Table~\ref{tab:AlternateFits}).
Jeng \cite{Jeng2007} gave equal weight to all measurement pairs (``P''), but this gives extreme weight to quantities with a large number ($N$) of measurements since the number of permutations grows as $(N-1)N/2$.
Giving each quantity equal weight (``Q'') also seems less fair, since a quantity measured many times will be weighted the same as a quantity measured only a few times.

\begin{table}[ht]
\newcommand{\DCBindent}{\hspace{5mm}}
\centering
\caption{\label{tab:AlternateFits}Fitted Student-t parameters for weighting by Quantities~(Q), Measurements (M, the default), Permutations (P), or using difference from weighted mean~(h).}
\medskip
\begin{tabular}{lccc}
\hline
				&$\nu$			& $\sigma$			& $\chi^2\textrm{/dof}$	\\
\hline
Interlab \\
\DCBindent Q	& $1.65\pm0.12$	& $1.35\pm	0.08$	&	$4.0$		\\
\DCBindent M	& $1.64\pm0.05$	& $1.62\pm	0.05$ 	&	$1.1$		\\
\DCBindent P	& $1.70\pm0.04$	& $1.76\pm	0.05$ 	&	$0.3$		\\
\DCBindent h	& $1.09\pm0.08$	& $2.06\pm	0.13$ 	&	$1.3$		\\
Nuclear \\
\DCBindent Q	& $1.93\pm0.07$	& $0.85\pm	0.02$ 	&	$1.8$		\\
\DCBindent M	& $1.99\pm0.06$	& $0.90\pm	0.02$ 	&	$1.6$		\\
\DCBindent P	& $2.19\pm0.07$	& $0.98\pm	0.03$ 	&	$1.4$		\\
\DCBindent h	& $1.82\pm0.06$	& $0.95\pm	0.02$ 	&	$1.4$		\\
Particle \\
\DCBindent Q	& $2.76\pm0.11$	& $1.01\pm	0.02$ 	&	$1.6$		\\
\DCBindent M	& $2.75\pm0.10$	& $1.05\pm	0.02$ 	&	$1.5$		\\
\DCBindent P	& $2.91\pm0.09$	& $1.14\pm	0.02$ 	&	$1.0$		\\
\DCBindent h	& $2.26\pm0.12$	& $1.10\pm	0.03$ 	&	$1.5$		\\
Medical \\
\DCBindent Q	& $3.44\pm0.16$	& $1.24\pm	0.02$ 	&	$1.2$		\\
\DCBindent M	& $3.30\pm0.11$	& $1.18\pm	0.02$ 	&   $0.7$		\\
\DCBindent P	& $3.59\pm0.12$	& $1.17\pm	0.03$ 	&   $0.4$		\\
\DCBindent h	& $3.00\pm0.18$	& $1.21\pm	0.04$ 	&	$0.8$		\\
\hline
\end{tabular}
\end{table}

Instead of using measurement pairs to study compatibility, Roos\,et\,al.\,\cite{Roos1975} instead calculated the weighted mean for each quantity, and then plotted the distribution of the uncertainty-normalized difference (``h'') from that mean for each measurement, i.e.
\begin{equation}
\label{eq:hDefinition}
h_{i}=\frac{|x_i - \bar{x}|}{\sqrt{u_i^2+u_{\bar{x}}^2}}
\end{equation}
where
\begin{equation}
\label{eq:weighting}
\bar{x}=\frac{\sum\limits_{i}{(x_i/u_i)^2}}{\sum\limits_{i}{1/u_i^2}} \qquad\text{and}\qquad \frac{1}{u_{\bar{x}}^2}=\sum\limits_{i}\frac{1}{u_i^2}
\end{equation}

$h$ is very similar to interlaboratory comparison $\zeta$-scores \cite{ISO13528}, which are the standard uncertainty normalized differences between measurements and an externally assigned value for the quantity.
The problem with using actual $\zeta$-scores is that they depend on having assigned values for the quantity independent of the measurements.
Such values are not usually available for the quantities studied here, so any assigned value must be determined from the measurements themselves, and such ``consensus values'' can be problematic \cite{ISO13528}.
The particular issue with $h$ is whether the weighted mean $\bar{x}$ is the best assigned value for a quantity given all the available measurements.
This is a reasonable assumption if the uncertainties are Normal, since then $\bar{x}$ from Equation~\ref{eq:weighting} is the maximum likelihood value for $x$ \cite{Roos1975}.
If the uncertainties are not Normal, however, $\bar{x}$ may be far from maximum likelihood, so it is not clear if $\bar{x}$ is the best choice for the assigned value.
Because of these issues, $z$ was preferred over $h$ in this study, but the $h$ and $z$ distributions are very similar.
As can be seen from Table~\ref{tab:AlternateFits}, the fit quality and parameter values are comparable for $h$ and $z$ distributions, except the tails appear even heavier in $h$.

\subsection{\label{ssec:Selected subsets}Selected data subsets}

To further investigate the variance in the distributions for different types of measurements, several additional data subsets were examined and their parameters listed in Table~\ref{tab:Special}.

\begin{table*}[ht]
\newcommand{\DCBindent}{\hspace{5mm}}
\centering
\caption{\label{tab:Special}Fitted Student-t parameters for selected data, with number of quantities, measurements, and comparison pairs.}
\medskip
\begin{tabular}{l|cc|ccc}
\hline
												&$\qquad\nu\qquad$	&$\,\qquad\sigma\qquad\quad\,$ &\quad Quant.	& \quad Meas.	& \quad Pairs \\
\hline
Key												&$1.9 \pm 0.1$		&$1.12\pm0.04$					& 372			& 3714			& 20308       \\
\DCBindent Key Metrology		            	&$3.2 \pm 0.2$		&$0.94\pm0.02$					& 197			& 2030			& 12070       \\
\DCBindent \DCBindent Selected	Metrology		&$9.9 \pm 2.6$		&$0.90\pm0.03$					& 156			&  575        	&   948 	  \\
\DCBindent Key Analytical		            	&$1.9 \pm 0.2$		&$1.62\pm0.13$					& 133  			& 1238          &  5938  	  \\
\DCBindent \DCBindent Selected	Analytical		&$2.1 \pm 0.3$		&$1.39\pm0.08$					& 127           &  503          &   848  	  \\
New Stable (since 2000)							&$7.6 \pm 1.3$		&$0.90\pm0.03$					& 357           & 1278          &  2478 	  \\
\DCBindent BABAR/BELLE Stable					&$6.7 \pm 2.1$	    &$0.91\pm0.04$					& 172  			&  435        	&   468 	  \\
\DCBindent Other New Stable						&$5.3 \pm 0.8$		&$0.79\pm0.03$					& 209  			&  752        	&  1395		  \\
Nuclear											&$2.0 \pm 0.1$		&$0.90\pm0.02$					&1437           &12380          & 66677  	  \\
\DCBindent Lifetimes							&$2.1 \pm 0.2$		&$1.30\pm0.09$					& 152  			& 1560          &  9779  	  \\
\DCBindent\DCBindent $u_x/x > 0.005$			&$2.2 \pm 0.2$		&$1.04\pm0.06$					& 125  			&  759          &  3123  	  \\
\DCBindent\DCBindent $u_x/x < 0.005$			&$2.8 \pm 0.5$		&$1.89\pm0.22$					& 110  			&  772          &  3503  	  \\
Constants										&$1.8 \pm 0.2$		&$0.89\pm0.06$					&   7           &  320        	&  9098  	  \\
\DCBindent Constants without G					&$3.2 \pm 0.5$		&$0.99\pm0.11$					&   6  			&  231        	&  5182  	  \\
\hline

\end{tabular}
\end{table*}

The Key Metrology data subset is for electrical, radioactivity, length, mass, and other similar physical metrology standards.
To see if the most experienced national laboratories were more consistent, Table~\ref{tab:Special} also lists Selected Metrology data from only the six national labs that reported the most Key Metrology measurements.
These laboratories were PTB (Physikalisch-Technische Bundesanstalt, Germany),  NMIJ (National Metrology Institute of Japan), NIST (National Institutes of Standards and Technology, USA), NPL (National Physical Laboratory, UK), NRC (National Research Council, Canada), and LNE (Laboratoire national de m\'etrologie et d'essais, France).
Similarly, Key Analytical chemistry data selected from the same national labs are also shown.
These are for measurements such as the amount of mercury in salmon, PCBs in sediment, or chromium in steel.
The metrology measurements by the selected national laboratories do have much lighter tails with $\nu\sim 10$, but this is not the case for their analytical measurements where $\nu\sim 2$.

New Stable particle data have the lightest tail in Table~\ref{tab:Fits}, but it is not clear if this is because the newer results have better determined uncertainties or are just more correlated.
The trend in particle physics is for fewer but larger experiments, and more than a third of the newer Stable measurements were made by just two very similar experiments (BELLE and BaBar), so the New Stable data is split into two groups in Table~\ref{tab:Special}.
There is no significant difference between the BELLE/Babar and Other experiments data.

Nuclear lifetimes with small and large relative uncertainties were compared.
They have similar tails, but the smaller uncertainty measurements appear to underestimate their uncertainty scales.

Measurements of Newton's gravitation constant are notoriously variable \cite{Speake2014,Faller2014}, so a data-set without $G_N$ results was examined.
The heavy tail is reduced, albeit with large uncertainty.

\subsection{Relative uncertainty}

\begin{figure}[ht]
\centering
\includegraphics[width=\figwidth]{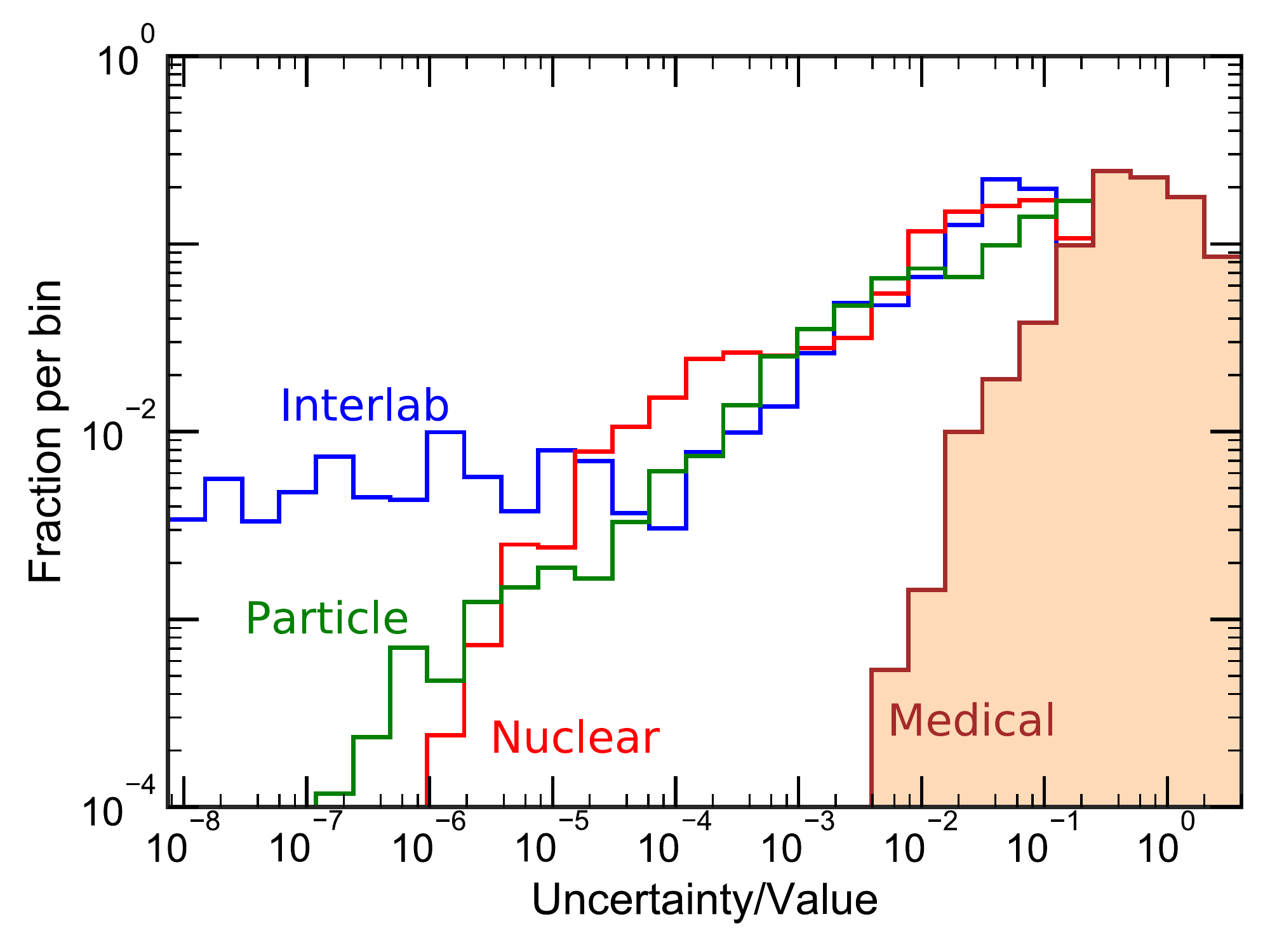}
\caption{\label{fig:Precision}Distribution of the relative uncertainty for data from {Fig.~\ref{fig:differential}}.}
\end{figure}

The accuracy of uncertainty evaluations appears to be similar in all fields, but unsurprisingly there are noticeable differences in the relative sizes of the uncertainties.
In particular, although individual physics measurements are not typically more reproducible than in medicine, they often have smaller relative uncertainty (i.e. uncertainty/value) as shown in Fig.~\ref{fig:Precision}.

\begin{figure}[ht]
\centering
\includegraphics[width=\figwidth]{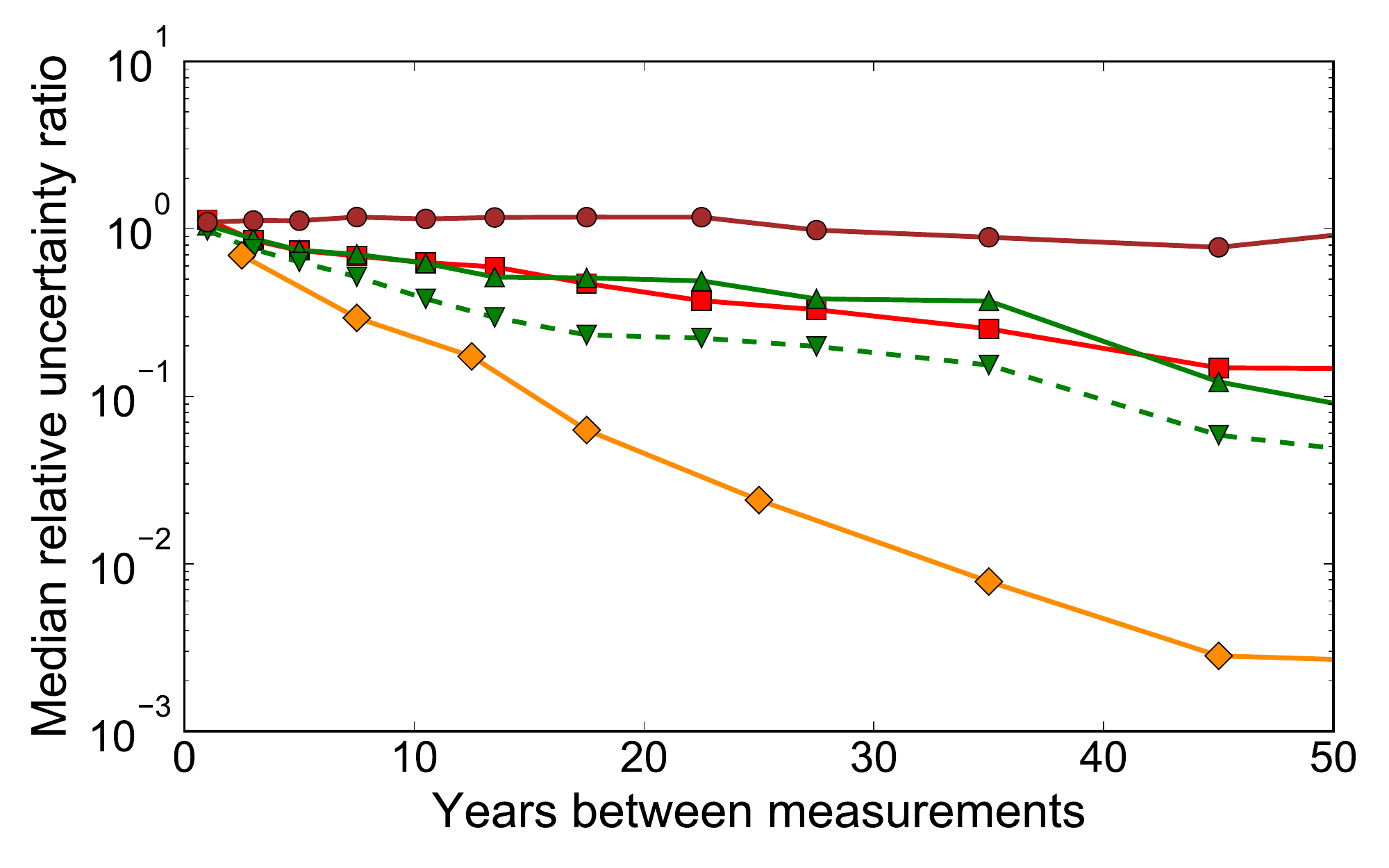}
\caption{\label{fig:PrecisionImprovement}Median ratio of the relative uncertainties (newer/older) for measurements in each $z$ pair as a function of the years between the two measurements:  Medical (brown circles), Particle (green triangles),  Nuclear (red squares), Stable (green dashed point-down triangles), Constants (orange diamonds).}
\end{figure}

Perhaps more importantly for discovery reproducibility, uncertainty improves more rapidly in physics than in medicine, as is shown in Fig.~\ref{fig:PrecisionImprovement}.
This difference in rates of improvement reflects the difference between measurements that depend on steadily evolving technology versus those using stable methods that are limited by sample sizes and heterogeneity \cite{Peterson2015}.
The expectation of reduced uncertainty in physics means that it is feasible to take a wait-and-see attitude towards new discoveries, since better measurements will quickly confirm or refute the new result.
Measurement uncertainty in Nuclear and Particle physics typically improves by about a factor of 2 every 15 years.
Constants data improve twice as fast, which is unsurprising since more effort is expected for more important quantities.

Physicists also tend not to make new measurements unless they are expected to be more accurate than previous measurements.
In the data sets reported here, the median improvement in uncertainty of Nuclear measurements compared to the best previous measurement of the same quantity is $u_{best}/u_{new} = 2.0\pm0.3$, and the improvement factors for Constants, Particle, and Stable measurements are $1.8\pm0.3$, $1.7\pm0.2$, and $1.3\pm0.1$.
In contrast, Medical measurements typically have greater uncertainties than the best previous measurements, with median $u_{best}/u_{new} =0.62\pm0.03$.
This is an understandable consequence of different uncertainty to cost relationships in physics and medicine.
Study population size is a major cost driver in medical research, so reducing the uncertainty by a factor of two can cost almost four times as much, which is rarely the case in physics.

\subsection{\label{ssec:Expectations}Expectations and correlations}

Prior expectations exist for most measurements reported here except for the Interlab data.
Such expectations may suppress heavy tails by discouraging publication of the anomalous results that populate the tails, since before publishing a result dramatically different from prior results or theoretical expectations, researchers are likely to make great efforts to ensure that they have not made a mistake.
Journal editors, referees and other readers also ask tough questions of such results, either preventing publication or inducing further investigation.
For example, initial claims \cite{OPERA2011,BICEP22014v1} of $6\sigma$ evidence for faster-than-light neutrinos and cosmic inflation  did not survive to actual publication \cite{OPERA2012,BICEP22014v3}.

\begin{figure}[ht]
\centering
\includegraphics[width=\figwidth]{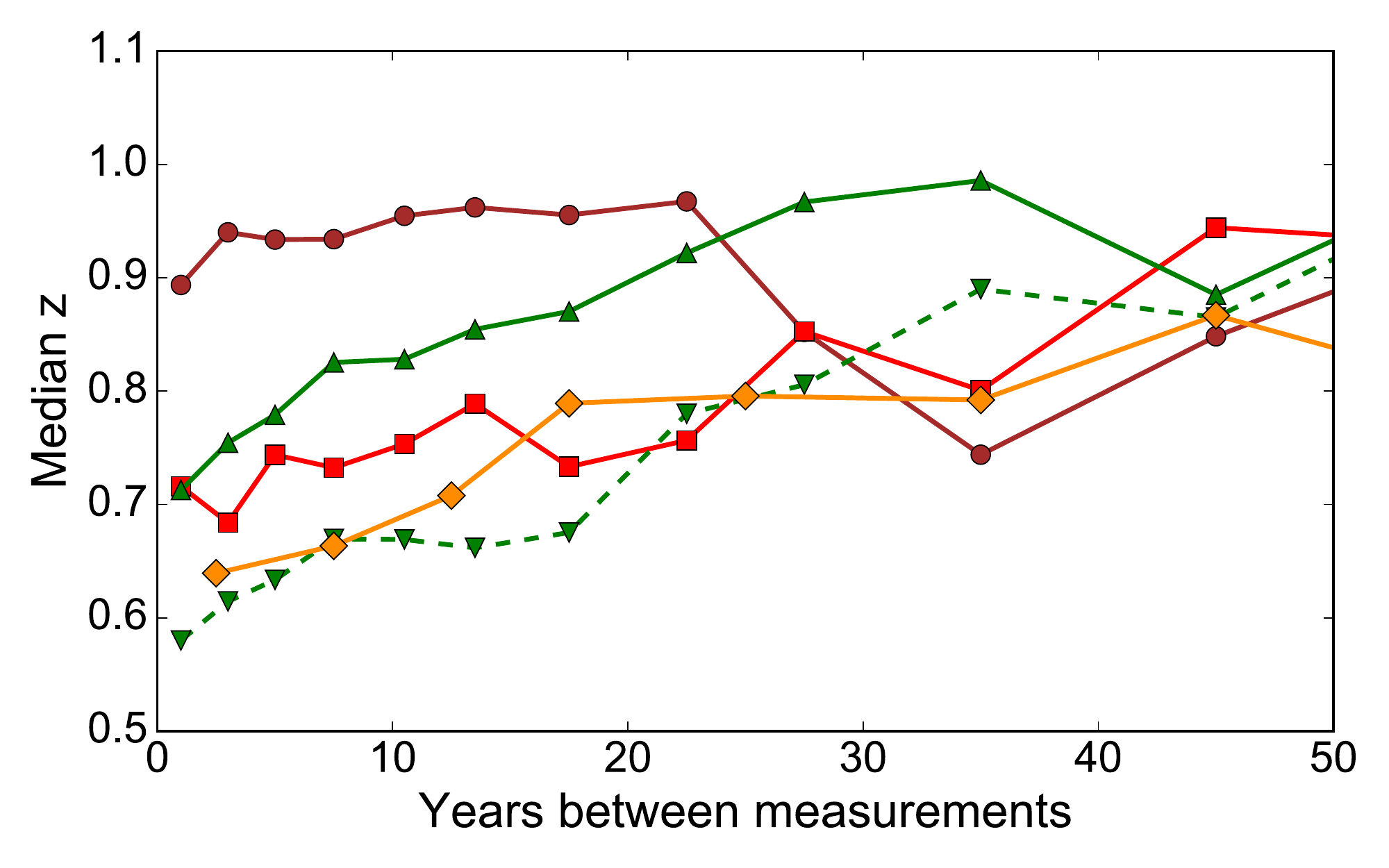}
\caption{\label{fig:correlation}Median $z$ value as a function of time difference between the two measurements in each $z$ pair: Medical (brown circles), Particle (green point-up triangles),  Nuclear (red squares),   Constants (orange diamonds), and  Stable (green dashed point-down triangles).}
\end{figure}

Fig.~\ref{fig:correlation} shows that Physics (Particle, Nuclear, Constants) measurements are more likely to agree if the difference in their publication dates is small.
Such ``bandwagon effects'' \cite{Jeng2007,Baker2013} are not observed in the Medical data, and they are irrelevant for Interlab quantities which are usually measured almost simultaneously.
These correlations imply that measurements are biased either by expectations or common methodologies.
Such correlations might explain the small ($<1$) values of $\sigma_x$ for Nuclear, Particle, and Constants data,  or it could be that researchers in these fields simply tend to overestimate the scale of their uncertainties \cite{Nachman2012}.
Removing expectation biases from the Physics data would likely make their tails heavier.

Although Interlab data are not supposed to have any expectation biases,  they are subject to methodological correlations due to common measurement models, procedures, and types of instrumentation, so even their tails would likely increase if all measurements could be made truly independent.

\Needspace{6\baselineskip} % Prevent bad page break here
\section{Discussion}\label{sec:Discussion}

\subsection{Comparison with earlier studies}

In a famous dispute with Cauchy in 1853, eminent statistician Ir\'{e}n\'{e}e-Jules Bienaym\'{e} ridiculed the idea that any sensible instrument had Cauchy uncertainties \cite{Bienayme1853}.
A century later, however,  Harold Jeffreys noted that systematic errors may have a significant Cauchy component, and that the scale of the uncertainty contributed by systematic effects depends on the size of the random errors \cite{Jeffreys1961}.

The results of this study agree with earlier research that also observed Student-t tails, but only looked at a handful of subatomic or astrophysics quantities up to $z \sim 5-10$~\cite{Roos1975,Chen2003,Hanson2007,Crandall2015Li,Crandall2015Dist}.
Unsurprisingly, the tails reported here are mostly heavier than those reported for repeated measurements made with the same instrument ($\nu\sim 3-9$) \cite{Jeffreys1938,Jeffreys1939,Dzhun2012}, which should be closer to Normal since they are not independent and share most systematic effects.

Instead of Student-t tails, exponential tails have been reported for several nuclear and particle physics data sets  \cite{Bukhvostov1973,Shlyakhter1994,Bukhvostov1997,Jeng2007}, but in all cases some measurements were excluded.
For example, the largest of these studies \cite{Jeng2007} looked at particle data (315 quantities, 53322 pairs) using essentially the same method as this paper, but rejected the 20\% of the data that gave the largest contributions to the $\chi^2$ for each quantity, suppressing the heaviest tails.
Despite this data selection, all these studies have supra-exponential heavy tails for $z\gtrsim5$, and so are qualitatively consistent with the results of this paper.
It is possible that averaging different quantities with exponential tails might produce apparent power-laws \cite{Anderson2001}, but this would require wild variations in the accuracy of the uncertainty estimates.

Instead of looking directly at the shapes of the measurement consistency distributions, Hedges \cite{Hedges1987} compared particle physics and psychology results and found them to have similar compatibility, with typically almost half of the quantities in both fields having statistically significant disagreements.

Thompson and Ellison reported substantial amounts of ``dark uncertainty'' in chemical analysis interlaboratory comparisons \cite{Thompson2011}.
Uncertainty is ``dark'' if it does not appear as part of the known contributions to the uncertainty of individual measurements, but is inferred to exist because the dispersion of measured values is greater than expected based on the reported uncertainties.
For example, six (21\%) of 28 BIPM Key Comparisons studied had ratios ($\bar{s}_{exp}/s_{obs}$) of expected to observed standard deviations less than 0.2.
This agrees with the Key Analytical results in Table~\ref{tab:Special} (which include some of the same Key Comparisons).
For sample sizes matching the 28 Comparisons, 20\% of samples drawn from a $\nu=2,\sigma=1.4$ Student-t distribution would be expected to have $\bar{s}_{exp}/s_{obs}<0.2$.
Pavese also noted the high rate of inconsistent Key Comparison measurements \cite{Pavese2015}.

The Open Science Collaboration (OSC) recently replicated 100 studies in psychology \cite{OSC2015}, providing some of the most direct evidence yet for poor scientific reproducibility.
Using the OSC study's supplementary information, $z$ can be calculated for 87 of the reported original/replication measurement pairs, and 27 (31\%) disagree by more than $2\sigma$, and 2 (2.3\%) by more than $5\sigma$.
This rate of disagreements is inconsistent with selection bias acting on a Normal distribution unless the $>\!5\sigma$ data are excluded, but can be explained by selection biased Student-t data with $\nu\sim 3$, consistent with the Medical data reported in Table~\ref{tab:Fits}.

\subsection{How measurements fail}
When a measurement turns out to be wrong, the reasons for this failure are often unknown, or at least unpublished, so it is interesting to look at examples where the causes were later understood or can be inferred.

For medical research, heterogeneity in methods or populations is a major source of variance.
The largest inconsistency in the Medical dataset is in a comparison of fever rates after acellular versus whole-cell pertussis vaccines \cite{CD001478}.
The large variance can likely be explained by significant differences among the study populations and especially in how minor adverse events were defined and reported.

The biggest $z$ values in the Particle data come from complicated multi-channel partial wave analyses of strong scattering processes, where many dozens of quantities (particle masses, widths, helicities, \ldots) are simultaneously determined.
Significant correlations often exist between the fitted values of the parameters but are not always clearly reported, and evaluations may not always include the often large uncertainties from choices in data and parameterization.

The largest disagreement in the Interlab data appears to be an obvious mistake.
In a comparison of radioactivity in water \cite{IAEA-AQ-15}, one lab reported an activity of $139352\pm0.82$ Bq/kg when the true value was about 31.
Even without knowing the expected activity, the unreasonably small fractional uncertainty should probably have flagged this result.
Such gross errors can produce almost-Cauchy deviations.
For example, if the numerical result of a measurement is simply considered as an infinite bit string, then any ``typographical'' glitch that randomly flips any bit with equal probability will produce deviations with a $1/x$ distribution.

One can hope that the best research will not be sloppy, but not even the most careful scientists can avoid all unpleasant surprises.
In 1996 a team from PTB (the National Metrological Institute of Germany) reported a measurement of $G_N$ that differed by $50\sigma$ from the accepted value; it took 8 years to track down the cause -- a plausible but erroneous assumption about their electrostatic torque transmitter unit \cite{Michaelis2004}.
A $6.5\sigma$ difference between the CODATA2006 and CODATA2010 fine structure constant values was due to a mistake in the calculation of some eighth-order terms in the theoretical value of the electron anomalous magnetic moment \cite{CODATA2010}.
A 1999 determination \cite{Fujii1999} of Avogadro's number by a team from Japan's National Research Laboratory of Metrology  using the newer x-ray crystal density method was off by $\sim9\sigma$ due to subtle silicon inhomogeneities \cite{DeBievre2001}.
In an interlaboratory comparison measuring PCB contamination in sediments, the initial measurement by BAM (the German Federal Institute for Materials Research and Testing) disagreed by many standard uncertainties, but this was later traced to cross-contamination in sample preparation \cite{CCQM-K25-2004}.
Several nuclear half-lives measured by the US National Institute for Standards and Technology were known for some years to be inconsistent with other measurements; it was finally discovered that a NIST sample positioning ring had been slowly slipping over 35 years of use \cite{Pomme2015}.

Often discrepancies are never understood and are simply replaced by newer results.
For example, despite bringing in a whole new research team to go over every component and system, the reason for a discordant NIST measurement of Planck's constant was never found, but newer measurements by the same group were not anomalous \cite{Gibney2015}.

\subsection{Causes of heavy tails}

Heavy tails have many potential causes, including bias \cite{Ioannidis2005}, overconfident uncertainty underestimates \cite{Henrion1986}, and uncertainty in the uncertainties \cite{Shlyakhter1994}, but it is not immediately obvious how these would produce the observed t distributions with so few degrees of freedom.

Even when the uncertainty $u$ is evaluated from the standard deviation of multiple measurements from a Normal distribution so that a Student-t distribution would be expected, there are typically so many measurements that $\nu$ should be much larger than what is observed.
Exceptions to this are when calibration uncertainties dominate, since often only a few independent calibration points are available, or when uncertainties from systematic effects are evaluated by making a few variations to the measurements, but these cannot explain most of the data.

Any reasonable publication bias applied to measurements with Gaussian uncertainties cannot create very heavy tails, just a distorted distribution with Gaussian tails -- to produce one false published $5\,\sigma$ result would require bias strong enough to reject millions of studies.
Underestimating $\sigma$ does not produce a heavy tail, only a broader Normal $z$ distribution.
Mixing multiple Normal distributions does not naturally produce almost-Cauchy distributions, except in special cases such as the ratio of two zero-mean Gaussians.

The heavy tails are not caused by poor older results.
The heaviest-tailed data in Fig.~\ref{fig:differential} are actually the newest -- 93\% of the interlaboratory data are less than 16 years old -- and eliminating older results taken prior to the year 2000 does not reduce the tails for most data as shown in Table~\ref{tab:Fits}.

Intentionally making up results, i.e. fraud, could certainly produce outliers, but this is unlikely to be a significant problem here.
Since most of the data were extracted from secondary meta-analyses (e.g. Review of Particle Properties, Table of Radionuclides, and Cochrane Systematic Reviews), results withdrawn for misconduct prior to the time of the review would likely be excluded.
One meta-analysis in the Medical dataset does include studies that were later shown to be fraudulent \cite{Carlisle2013}, but the fraudulent results actually contribute slightly less than average to the overall variance among the results for that meta-analysis.

\subsection{\label{ssec:Modelling}Modelling}
Modelling the heavy tails may help us understand the observed distributions.
One way is to assume that the measurement values are normally distributed with standard deviation $t$ that is unknown but which has a probability distribution $f(t)$ \cite{Bukhvostov1973,Shlyakhter1994,Bukhvostov1997,Dose1999,Hanson2007}.
The measured value $x$ is then expected to have a probability distribution
\begin{equation}
\label{eq:Shlyakhter}
\mathcal{P}(x)=\int_0^\infty dt f\!(t)\frac{1}{\sqrt{2\pi}t}e^{-x^2/(2t^2)}.
\end{equation}
This is essentially a Bayesian estimate with prior $f\!(t)$ and a Normal likelihood with unknown variance.
If the uncertainties are accurately evaluated and Normal with variance $\sigma^2$, $f\!(t)$ will be a narrow peak at $t=\sigma$.
Assuming that $f\!(t)$ is a broad Normal distribution leads to exponential tails~\cite{Shlyakhter1994} for large $z$.

In order to generate Student-t distributions, $f\!(t)$ must be a scaled inverse chi-squared (or Gamma) distribution in $t^2$ \cite{Dose1999,Hanson2007}.
This works mathematically, but why would variations in $\sigma$ for independent measurements have such a distribution?

Heavy tails can only be generated by effects that can produce a wide range of variance, so we must model how consistency testing is used by researchers to constrain such effects.
Consistency is typically tested using a metric such as the calculated chi-squared statistic for the agreement of $N$ measurements $x_i$ \cite{BohmZech2010}
\begin{equation}
\chi^2_c(x,u)  = \sum\limits_{i=1}^N \frac{\left ( x_i - \bar{x} \right )^2}{u_i^2}
\end{equation}
where $\bar{x}$ is the $x_i$ weighted mean and $u_i$ are the standard uncertainties reported by the researchers.
For accurate standard uncertainties, $\chi^2_c$ will have a chi-squared probability distribution
with \mbox{$\nu\!\!=\!\!N\!\!-\!\!1$}.
If, however, the reported uncertainties are incorrect and the true standard uncertainties are $t u_i$, then it will be $\chi^2_{true}(x,t u) = \chi^2_c(x,u)/t^2$ that is chi-squared distributed.

Researchers will likely search for problems if different consistency measurements have a poor $\chi^2_c(x,u)$, which typically means  $\chi^2_c(x,u) > \nu$.
The larger an unknown systematic error is, the more likely it is to be detected and either corrected or included in the reported uncertainty, so published results typically  have $\chi^2_c(x,u) \sim \nu$.
Since $\chi^2_c(x,u)/t^2$ is expected to have a chi-squared distribution, a natural prior for $t^2$ is indeed the scaled inverse chi-squared distribution needed to generate Student-t distributions from Equation~\ref{eq:Shlyakhter}.

More mechanistically, it could be assumed that a Normally distributed systematic error will be missed by $N_m$ independent measurements if their $\chi^2(u)=(t^2/u^2)\chi^2(t)$ is less than some threshold $\chi^2_{max}\sim \nu = N_m-1$.
If the distribution of all possible systematic effects is $P_0(t)$, then the probability distribution for the unfound errors will be
\begin{equation}
\label{eq:model}
f\!\left(t;\nu \right)  =  P_0 \left ( t \right) F(\chi^2_{max}/t^2 ;  \nu)
\end{equation}
where $F$ is the cumulative $\chi^2$ distribution.
$P_0(t)$ is unknown, but a common Bayesian scale-invariant choice is $P_0(t) \propto 1/t^\alpha$, with ${\alpha>0}$.

Using this model with the reported uncertainty $\sigma$ as the lower integration bound, the curve generated from Equations~\ref{eq:Shlyakhter} and \ref{eq:model} is very close to a $\nu=N_m-1+\alpha$ Student-t distribution.
The observed small values for $\nu$ mean that both $N_m$ and $\alpha$ must be small.
Making truly independent consistency tests is difficult, so it is not surprising that the effective number of checks ($N_m$) is usually small.

This model is plausible, but why are systematic effects consistent with a $P_0(t) \propto 1/t^\alpha$  power-law size distribution?

\subsection{\label{ssec:Complex systems}Complex systems}

Scientific measurements are made by complex systems of people and procedures, hardware and software, so one would expect the distribution of scientific errors to be similar to those produced by other comparable systems.

Power-law behaviour is ubiquitous in complex systems~\cite{Clauset2009}, with the cumulative distributions of observed sizes ($s$) for many effects falling as $1/s^\alpha$, and these heavy tails exist even when the system has been designed and refined for optimal results.

A Student-t distribution has cumulative tail exponent $\alpha=\nu$, and the values for $\nu$ reported here are consistent with power-law tails observed in other designed complex systems.
The frequency of software errors typically has a cumulative power-law tail corresponding to small $\nu\sim2-3$ \cite{Hatton2012}, and in scientific computing these errors can lead to quantitative discrepancies orders of magnitude greater than expected~\cite{Hatton1994}.
The size distribution of electrical power grid failures has $\nu\sim 1.5-2$ \cite{Dobson2007}, and the frequency of spacecraft failures has $\nu\sim 0.6-1.3$ \cite{Karimova2011}.
Even when designers and operators really, really want to avoid mistakes, they still occur: the severity of nuclear accidents falls off only as $\nu\sim 0.7$~\cite{Sornette2013}, similar to the power-laws observed for the sizes of industrial accidents \cite{Lopes2015} and oil spills~\cite{Englehardt2002}.
Some complex medical interventions have power-law distributed outcomes with $\nu\sim 3-4$ \cite{Burton2012}.

Combining the observed power-law responses of complex systems with the power-law constraints of consistency checking for systematic effects discussed in Section \ref{ssec:Modelling}, leads naturally to the observed consistency distributions with heavy power-law tails.
There are also several theoretical arguments that such distributions should be expected.

A systematic error or mistake is an example of a risk analysis incident, and power-law distributions are the maximal entropy solutions for such incidents when there are multiple nonlinear interdependent causes~\cite{Englehardt2002}, which is often the case when things go wrong in research.

Scientists want to make the best measurements possible with the limited resources they have available, so scientific research endeavours are good examples of highly structured complex systems designed to optimize outcomes in the presence of constraints.
Such systems are expected to exhibit ``highly optimized tolerance'' \cite{Carlson1999,Carlson2002}, being very robust against designed-for uncertainties, but also hypersensitive to unanticipated effects, resulting in power-law distributed responses.
Simple continuous models for highly optimized tolerant systems are consistent with the heavy tails observed in this study.
These models predict that $\alpha\sim1+1/d$ \cite{Newman2002,Carlson2002}, where $d(>0)$ is the effective dimensionality of the system, but larger values of $\alpha$ arise when some of the resources are used to avoid large deviations \cite{Newman2002}, e.g. spending time doing consistency checks.

\subsection{How can heavy tails be reduced?}\label{ssec:Reduce Tails}

If one believes that mistakes can be eliminated and all systematic errors found if we just work hard enough and apply the most rigorous methodological and statistical techniques, then results from the best scientists should not have heavy tails.
Such a belief, however, is not consistent with the experienced challenges of experimental science, which are usually hidden in most papers reporting scientific measurements \cite{Collins2001,Franklin2013}.
As Beveridge famously noted \cite{Beveridge1957}, often everyone else believes an experiment more than the experimenters themselves.
Researchers always fear that there are unknown problems with their work, and traditional error analysis cannot ``include what was not thought of'' \cite{Faller2014}.

It is not easy to make accurate \textit{a priori} identifications of those measurements that are so well done that they avoid having almost-Cauchy tails.
Expert judgement is subject to well-known biases \cite{Sutherland2015}, and obvious criteria to identify better measurements may not work.
For example, the Open Science Collaboration found that researchers' experience or expertise did not significantly correlate with the reproducibility of their results \cite{OSC2015} -- the best predictive factor was simply the statistical significance of the original result.
The best researchers may be better at identifying problems and not making mistakes, but they also tend to choose the most difficult challenges that provide the most opportunities to go wrong.

Reducing heavy tails is challenging because complex systems exhibit scale invariant behaviour such that reducing the size of failures does not significantly change the shape of their distribution.
Improving sensitivity makes previously unknown small systematic issues visible so they can be corrected or included in the total uncertainty.
This improvement reduces $\sigma$, but even smaller systematic effects now become significant and tails may even become heavier and $\nu$ smaller.
Comparing Figures~\ref{fig:differential} and \ref{fig:Precision}, it appears that data with higher average relative uncertainty tend to have heavier tails.
This relationship between relative uncertainty and measurement dispersion is reminiscent of the empirical Horwitz power-law in analytical chemistry \cite{Horwitz2006}, where the relative spread of interlaboratory measurements increases as the required sensitivity gets smaller, and of Taylor's Law in ecology where the variance grows with sample size so that the uncertainty on the mean does not shrink as $1/\sqrt{N}$ \cite{Eisler2008}.

In principle, statistical errors can be made arbitrarily small by taking enough data, and $\nu$ can be made arbitrarily large by making enough independent consistency checks, but researchers have only finite time and resources so choices must be made.
Taking more consistency check data limits the statistical uncertainty, since it is risky to treat data taken under different conditions as a single data set.
Consistency checks are never completely independent since it is impossible for different measurements of the same quantity not to share any people, methods, apparatus, theory or biases, so researchers must decide what tests are reasonable.
The observed similar small values for $\nu$ may reflect similar spontaneous and often unconscious cost-benefit analyses made by researchers.

The data showing the lightest tail reported here (in Table~\ref{tab:Special}) may provide some guidance and caution.
The high quality of the Selected Metrology standards measurements by leading national laboratories shows that heavy tails can be reduced by collaboratively taking great care to ensure consistency by sharing methodology and making regular comparisons.
There are, however, limits to what can be achieved, as illustrated by the much heavier tail of the analytical standards measured by the same leading labs.
Secondly, consistency is easier than accuracy.
Interlaboratory comparisons typically take place over relatively short periods of time, with participating institutions using the best standard methods available at that time.
Biases in the standard methods may only be later discovered when new methods are introduced.
For example, work towards a redefinition of the kilogram and the associated development of new silicon atom counting technology revealed inconsistencies with earlier watt-balance measurements, and this has driven improvements in both methods \cite{Gibney2015}.
Finally, selection bias that hides anomalous results is hard to eliminate.
For one metrology key comparison, results from one quantity were not published because some laboratories reported ``incorrect results'' \cite{CCL-K1}.

Reducing tails is particularly challenging for measurements where the primary goal is improved sensitivity that may lead to new scientific understanding.
By definition, a better measurement is not an identical measurement, and every difference provides room for new systematic errors, and every improvement that reduces the uncertainty makes smaller systematic effects more significant.
Frontier measurements are always likely to have heavier tails.

\section{Conclusions}\label{sec:Conclusions}

Published scientific measurements typically have non-Gaussian almost-Cauchy $\nu \sim\,2-4$ \mbox{Student-t} error distributions, with up to $10\%$ of results in disagreement by $>\!5\sigma$.
These heavy tails occur in even the most careful modern research,
and do not appear to be caused by selection bias, old inaccurate data, or sloppy measurements of uninteresting quantities.
For even the best scientists working on well understood measurements using similar methodology, it appears difficult to achieve consistency better than $\nu \sim 10$, with about $0.1\%$ of results expected to be $>\!5\sigma$ outliers, a rate a thousand times higher than for a Normal distribution.
These may, however, be underestimates.
Because of selection/confirmation biases and methodological correlations, historical consistency can only set lower bounds on heavy tails --  multiple measurements may all agree but all be (somewhat) wrong.

The effects of unknown systematic problems are not completely unpredictable.
Scientific measurement is a complex process and the observed distributions are consistent with unknown systematics following the low-exponent power-laws that are theoretically expected and experimentally observed for fluctuations and failures in almost all complex systems.

Researchers do determine the scale of their uncertainties with fair accuracy, with the scale of Medical uncertainties ($\sigma_{x} \sim 1$) slightly more consistent with the expected value ($\sigma_{x} = 1$) than in Physics ($\sigma_{x} \sim 0.7-0.8$).
Medical and Physics research have comparable reproducibility in terms of how well different studies agree within their uncertainties, consistent with a previous comparison of particle physics with social sciences \cite{Hedges1987}.
Medical research may have slightly lighter tails, while Physics results typically have better relative uncertainty and greater statistical significance.

Understanding that error distributions are often almost-Cauchy should encourage use of t-based \cite{Lange1989}, median \cite{Gott2001}, and other robust statistical methods \cite{Medina2015}, and supports choosing Student-t \cite{Gelman2006} or Cauchy \cite{Polson2012} priors in Bayesian analysis.
Outlier-tolerant methods are already common in modern meta-analysis, so there should be little effect on accepted values of quantities with multiple published measurements, but this better understanding of the uncertainty may help improve methods and encourage consistency.

False discoveries are more likely if researchers apply Normal conventions to almost-Cauchy data.
Although much abused, the historically common use of $p<0.05$ as a discovery criterion suggests that many scientists would like to be wrong less than 5\% of the time.
If so, the results reported here support the nominal 5-sigma discovery rule in particle physics, and may help discussion of more rigorous significance criteria in other fields \cite{Johnson2013,Gelman2014,Colquhoun2014}.

This study should help researchers better understand the uncertainties in their measurements, and may help decision makers and the public better interpret the implications of scientific research \cite{Fischhoff2014}.
If nothing else, it should also remind everyone to never use Normal/Gaussian statistics when discussing the likelihood of extreme results.

\section*{Acknowledgements}

I thank the students of the University of Toronto Advanced Undergraduate Physics Lab for inspiring consideration of realistic experimental expectations, and the University of Auckland Physics Department for their hospitality during very early stages of this work.
I am grateful to D. Pitman for patient and extensive feedback, to R. Bailey for his constructive criticism, to R. Cousins, D. Harrison, J. Rosenthal, and P. Sinervo  for useful suggestions and discussion, and to M. Cox for many helpful comments on the manuscript.

\section*{Data accessibility}
The sources for all data analysed are listed in the associated ancillary file: UncertaintyDataDescription.xls.

\raggedright

\onecolumngrid

\end{document}